# Extracting lung function-correlated information from CT-encoded static textures

Yu-Hua Huang[1], Xinzhi Teng[1], Jiang Zhang[1], Zhi Chen[1], Zongrui Ma[1], Ge Ren[1], Feng-Ming (Spring) Kong[2], Jing Cai[1,†]

[1] Department of Health Technology and Informatics, The Hong Kong Polytechnic University, Hong Kong SAR

[2] Department of Clinical Oncology, The University of Hong Kong, Hong Kong SAR

[†]**Corresponding Author:** Jing Cai, Ph.D.　Email: jing.cai@polyu.edu.hk

## Abstract

**Purpose:**

Breathing-induced surrogate changes are commonly used to assess lung function from CT. However, the inherent characteristics of lung tissues, which are independent of breathing manoeuvre, may provide more fundamental information on lung function. This paper attempted to study function-correlated lung textures and their spatial distribution from CT using a sparse-to-fine radiomics framework.

**Materials and Methods:**

Twenty-one lung cancer patients with thoracic 4DCT scans, DTPA-SPECT ventilation images $V_{NM}$, and available pulmonary function test (PFT) measurements were collected. 79 radiomic features were included for analysis, and a sparse-to-fine strategy including subregional feature discovery and voxel-wise feature distribution study was carried out to identify the function-correlated radiomic features. At the subregion level, lung CT images were partitioned and labeled as defected/non-defected patches according to reference $V_{NM}$. At the voxel-wise level, feature maps (FMs) of selected feature candidates were generated for each 4DCT phase. Quantitative metrics, including Spearman coefficient of correlation (SCC) and Dice similarity coefficient (DSC) for FM-$V_{NM}$ spatial agreement assessments, intra-class coefficient of correlation (ICC) for FM inter-phase robustness evaluations, and FM-PFT comparisons, were applied to validate the results.

**Results:**

At the subregion level, eight function-correlated features were filtered out with medium-to-large statistical strength (effect size >0.330) to differentiate defected/non-defected lung regions.

At the voxel-wise level, FMs of candidates yielded moderate-to-strong voxel-wise correlations with reference $V_{NM}$. Among them, FMs of GLDM Dependence Non-uniformity showed the highest robust (ICC=0.96, $p$<0.0001) spatial correlation, with median SCCs ranging from 0.54 to 0.59 throughout ten breathing phases. Its phase-averaged FM achieved a median SCC of 0.60, the median DSC of 0.60/0.65 for high/low functional lung volumes, respectively, and the correlation of 0.646 between the spatially averaged feature values and PFT measurements.

**Conclusion:**

Candidate function-correlated features with their corresponding FMs were discovered and analyzed. The results hold the exciting potential to further the understanding of the underlying correlation between regional function and lung textural information, and to facilitate more accurate lung disease diagnosis. Validations on the generalizability of the FMs and standardization of implementation protocols are warranted in the future before further clinically relevant investigations.

**Keywords:** CT, ventilation, lung cancer, functional imaging, radiomics

## 1. Introduction

Spatial variability in the distribution of lung function, of which ventilation (V) and perfusion (Q) are fundamental components, can be commonly affected by the mixture of internal and external, physiological and pathological factors [1][2][3]. Visual inspections and quantitative assessments of the lung function distribution have unique applications for interpretation of respiratory mechanisms, diagnosis of thoracic diseases, and for personalized planning and lung toxicity optimization in lung cancer radiotherapy (RT) planning (e.g., functional lung avoidance radiotherapy, FLART) [4][5].

At present, the mainstream methods for lung function imaging depend on contrast agent-based direct mapping or algorithm-aided indirect mapping. Nuclear medicine (NM) imaging modalities are widely used through measuring the concentration of radiotracers, including single photon emission computed tomography (SPECT) [6], and positron emission tomography (PET) [7]. The experimental contrast-enhanced magnetic resonance imaging (CE MRI) and computed tomography (CT) techniques are also being tested [8][9][10]. However, NM-based function imaging commonly provides limited quality and radiotracer clumping artifacts. CE MRI/CT is challenging to implement clinically due to either the limited availability of contract agents or the complicated procedure. In addition to these modalities that can directly map lung function, other algorithm-aided techniques have been developed to obtain functional surrogates by adding pre- or post-processing to the standard imaging process, including CT-derived ventilation/perfusion imaging (CTVI/PI) [11][12], MRI spin labeling [13][14][15], and Fourier-decomposition V/Q MRI [16]. Among them, CTVI/PI is highly attractive, and a series of metrics have been proposed over the past two decades. Deformable image registration (DIR) based CTVI/PI methods can mathematically derive function distribution via the registered biphasic breath-hold CT images (BHCT), or the biphasic image pair from the four-dimensional computed tomography (4DCT) scans. Few artificial intelligence (AI-) based CTVI/PI models [17][18][19] were also proposed to directly generate NM-alike lung function images through the end-to-end convolutional neural network (CNN). However, DIR-based CTVI/PI has been demonstrated to be sensitive to the input image quality and DIR algorithm. The limited accuracy and robustness provided by DIR-based CTVI/PI still fail to meet the expectations of clinical implementations. Existing applications of AI-based CTVI/PI are still in an early stage. The automatic data-driven image feature extraction and learning process did not consider the physiological and pathological mechanism of lung function, rendering insufficient model explainability and reliability.

The applications of quantitative texture analysis in thoracic CT images (i.e., radiomics) [20][21] offer a different perspective for studying lung function based on the assumption that the inherent characteristics of pulmonary tissues, which are independent of breathing manoeuvre, may provide more fundamental information on lung physiology and pathology [22][23]. At the same time, the extraction of functional information is expected to be more explainable and robust due to the mathematical quantification and standardization of various image textural features. Therefore, in this study, we aimed to extract and identify underlying lung function-correlated textures, and to study their spatial distribution and robustness from thoracic CT images, relying on the well-defined radiomic feature categories using a sparse-to-fine strategy.

## 2.  Methods

## 2.1  Overall study design

Figure 1 shows the overall study workflow to screen and analyze the underlying function-correlated radiomics features. A sparse-to-fine strategy including subregional feature discovery and voxel-wise feature distribution study was used in this research for fast and comprehensive lung function-correlated feature discovery. In the subregional feature discovery stage, time-average 4DCT images of all patients were firstly divided into patches. According to corresponding reference $V_{NM}$, each CT image patch was binarily labeled as defected or non-defected. Within each defected or non-defected patch, the subregional radiomic feature were comprehensively calculated to extract the intensity and texture information. Statistical analysis was performed to identify limited numbers of function-correlated feature candidates for the following study. The feature maps (FMs) of selected feature candidates were generated using the voxel-wise radiomics approach for all patients. Quantitative metrics, including Spearman coefficient of correlation (SCC) and Dice similarity coefficient (DSC) for FM-$V_{NM}$ spatial agreement assessments, intra-class coefficient of correlation (ICC) for FM inter-phase robustness evaluations, and FM-PFT comparisons, were applied to validate the results.

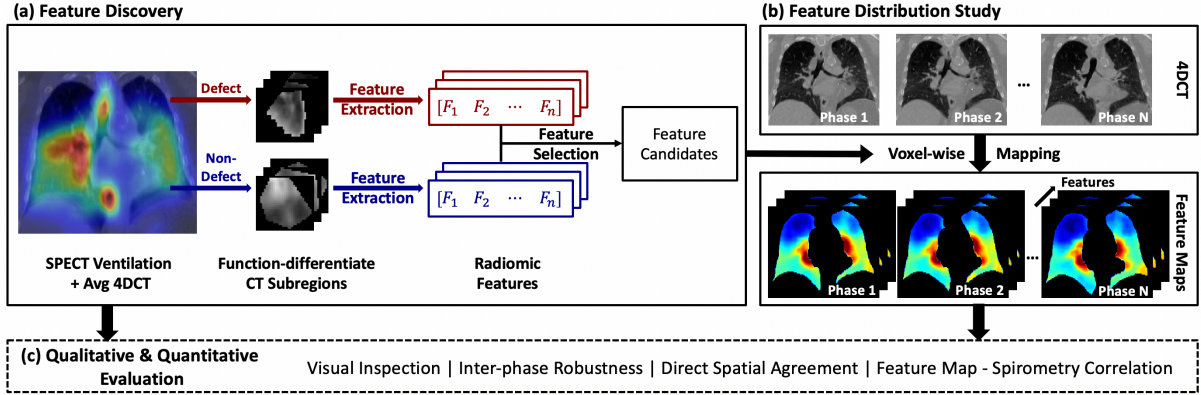

**Figure 1** The designed workflow of this study. Avg = average

## 2.2 Dataset and image preprocessing

A total number of 21 lung cancer RT patients from the public dataset of VAMPIRE Challenge [11] were retrospectively collected for this study. Free-breathing ten-phase 4DCT scan was acquired for each patient for RT treatment planning, with reconstructed resolutions between $0.97 \times 0.97 \times 2$ mm$^3$ to $0.97 \times 0.97 \times 3$ mm$^3$. The patients also underwent DTPA-SPECT scans for ventilation assessment. And for each patient, the SPECT image was rigidly registered and interpolated to link the spatial position and dimension with 4DCT images. Besides, standard pulmonary function test (PFT) measurements were also available for 13 patients of them. Spirometry tests including the forced expiratory volume in one second (FEV1, % predicted), the forced vital capacity (FVC, % predicted), and the ratio of FEV1/FVC (%) were measured according to the standard protocol from the American Thoracic Society and the European Respiratory Society [24].

All images were resampled to an identical isotropic voxel size of $1.0 \times 1.0 \times 1.0$ mm$^3$ using the B-spline interpolator. In order to reduce the computation scale of the subsequent steps without losing the information within the lung regions, for each patient, all their images (4DCT phases, time-average 4DCT, and SPECT ventilation scans) were cropped to the minimum permissible dimension defined by the pulmonary motion envelope.

## 2.3 Subregional feature discovery

### 2.3.1 Patch generation and defect classification

Masked Simple Linear Iterative Clustering (*mask*SLIC) algorithm [25][26] was used to generate compact and quasi-equally sized subregions within each patient's lung. The resulting segmentations were subsequently applied to the corresponding time-average 4DCT and SPECT ventilation images to partition them into patches.

With the guidance of reference $V_{NM}$ image, the CT image patches could be accordingly classified into defected and non-defected patches by comparing the mean $V_{NM}$ signal within the patch ROIs with predetermined patient-specific thresholds. The definition of the signal threshold between defect and non-defect in this study referred to a metric developed based on precedent from nuclear medicine ventilation assessment. In brief, this metric distinguished lung regions that deviate more than 15% from a hypothetically homogeneous ventilation distribution in the normal lung [27][28].

### 2.3.2 Subregional feature extraction and redundancy analysis

Within each defected or non-defected lung patch, the subregional radiomic feature values were extracted using Radiotherapy Data Analysis and Reporting (RADAR) tool, our in-house platform for quantitative radiomics analysis. The grey levels of input time-average 4DCT images were discretized into 64 bins within the range of [-1000, 0] HU prior to subregional feature extraction. In this study, we included 11 first-order intensity-based features as well as 68 second-order texture-based features from the original image domain, which can be classified into four feature categories: (a) grey level co-occurrence matrix (GLCM) features, (b) grey level size zone matrix (GLSZM) features, (c) grey level run length matrix (GLRLM) features, and (d) grey level dependence matrix (GLDM) features. The full name list of the features used in this study is provided in Table 1.

To investigate the redundancy of features and also to ensure that redundant features will not be simultaneously taken into account in the subsequent usage of features, the Density-Based Spatial Clustering of Applications with Noise (DBSCAN) clustering [29] with correlation-based pair-wise distance measurement was performed to identify highly correlated feature groups. Specifically, the minimum allowed correlation between features was set as 0.90 for one to be assigned in the same redundant subset of the other.

**Table 1** The completed radiomics feature name list.

| Source Image | Feature Class | Feature Name |
|---|---|---|
| Original Image | First Order | Energy |
| | | Entropy |
| | | Kurtosis |
| | | Mean |
| | | Mean Absolute Deviation |
| | | Robust Mean Absolute Deviation |
| | | Root Mean Squared |
| | | Skewness |
| | | Total Energy |
| | | Uniformity |
| | | Variance |
| | Grey Level Co-occurrence Matrix (GLCM) | Autocorrelation |
| | | Cluster Prominence |
| | | Cluster Shade |
| | | Cluster Tendency |
| | | Contrast |
| | | Correlation |
| | | DifferenceAverage |
| | | DifferenceEntropy |
| | | DifferenceVariance |

| | |
|---|---|
| | Inverse Difference |
| | Inverse Difference Moment |
| | Inverse Difference Moment Normalized |
| | Inverse Difference Normalized |
| | Informational Measure of Correlation 1 |
| | Informational Measure of Correlation 2 |
| | InverseVariance |
| | JointEnergy |
| | JointEntropy |
| | MaximumProbability |
| | SumAverage |
| | SumEntropy |
| | SumSquares |
| **Grey Level Dependence Matrix (GLDM)** | DependenceEntropy |
| | DependenceNonUniformity |
| | DependenceNonUniformityNormalized |
| | DependenceVariance |
| | GrayLevelNonUniformity |
| | GrayLevelVariance |
| | HighGrayLevelEmphasis |
| | LargeDependenceEmphasis |
| | LargeDependenceHighGrayLevelEmphasis |
| | LargeDependenceLowGrayLevelEmphasis |
| | LowGrayLevelEmphasis |
| | SmallDependenceEmphasis |
| | SmallDependenceHighGrayLevelEmphasis |
| | SmallDependenceLowGrayLevelEmphasis |
| **Grey Level Run Length Matrix (GLRLM)** | GrayLevelNonUniformity |
| | GrayLevelNonUniformityNormalized |
| | GrayLevelVariance |
| | HighGrayLevelRunEmphasis |
| | LongRunEmphasis |
| | LongRunHighGrayLevelEmphasis |
| | LongRunLowGrayLevelEmphasis |
| | LowGrayLevelRunEmphasis |
| | RunEntropy |
| | RunLengthNonUniformity |
| | RunLengthNonUniformityNormalized |
| | RunPercentage |
| | RunVariance |
| | ShortRunEmphasis |
| | ShortRunHighGrayLevelEmphasis |
| | ShortRunLowGrayLevelEmphasis |
| **Grey Level Size Zone Matrix (GLSZM)** | GrayLevelNonUniformity |
| | GrayLevelNonUniformityNormalized |
| | GrayLevelVariance |
| | HighGrayLevelZoneEmphasis |
| | LargeAreaEmphasis |
| | LargeAreaHighGrayLevelEmphasis |

LargeAreaLowGrayLevelEmphasis
LowGrayLevelZoneEmphasis
SizeZoneNonUniformity
SizeZoneNonUniformityNormalized
SmallAreaEmphasis
SmallAreaHighGrayLevelEmphasis
SmallAreaLowGrayLevelEmphasis
ZoneEntropy
ZonePercentage
ZoneVariance

### 2.3.3 Subregional feature candidate selection

In this step, feature selection is performed to filter out a limited number of function-correlated radiomic features at the subregional level. These potential feature candidates are expected to have statistically significant and sufficiently large expression differences between previous-defined defected and non-defected lung CT patches in the patient cohort.

Iterate through each patient, we performed the non-parametric Mann-Whitney U test for each feature's value in defected and non-defected lung CT patches. In addition to the frequently used p-values, we also calculated the effect size (ES) for each test. ES is a unitless value evaluating the strength of a statistical claim. In simple terms, by calculating the ES of a statistical test, it can tell us how much difference there is between the distributions of two sets of statistics. As a non-parametric rank-sum test, the rank-biserial correlation (RBC) in the two-sample case [30] was used for assessing the ES for the Mann-Whitney U test.

Figure 2 shows the statistical analysis workflow of the feature selection task. For every single feature, the Mann-Whitney U test-derived ES for all patients were averaged. Thus, one single value (the mean ES) would be obtained for each radiomic feature as the index to statistically measure its ability to distinguish between defected or non-defected CT patches.

The 79 features listed in the previous section were sorted in descending order by the mean ES to filter out potential function-correlated feature candidates.

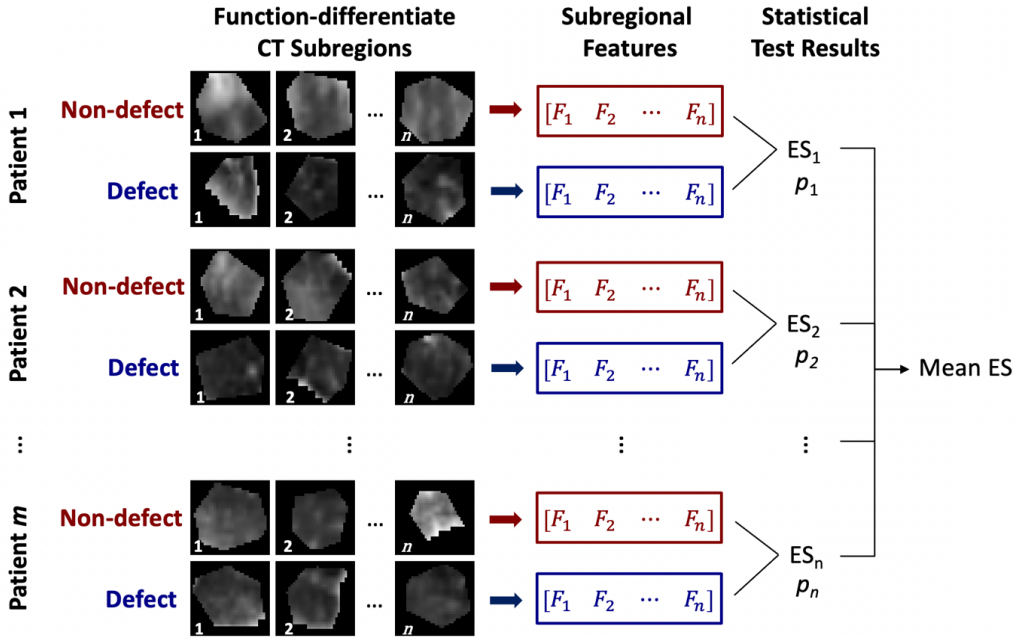

**Figure 2** Schematic for the statistical analysis workflow of a single radiomic feature. One single value (the mean ES) will be provided for each feature to statistically measure its ability to distinguish between defected or non-defected CT patches. ES = effect size.

## 2.4 Voxel-wise feature distribution study

We calculated the 3D FMs for the before-selected feature candidates to study the spatial texture distribution. As shown in Figure 1, for each patient's 4DCT phase image, a 3D sliding window with the size of [21, 21, 21] mm was applied to voxel-by-voxel map the spatially encoded regional response of the before-selected eight radiomic features. For a particular lung voxel, such a sliding window could extract the local intensity and textural characteristics in its neighborhood. To avoid introducing non-lung tissues, voxels outside the lung mask were excluded from the calculation (i.e., the mask-preserved sliding window). The FM computing extension of RADAR used in this study was fully optimized for parallel computing to accelerate computation further.

Similar to the image preprocessing in subregional feature extraction, the grey levels of each input 4DCT phase image were discretized into 64 bins within the range of [-1000, 0] HU prior to FM calculation. This process generated ten phase-specific FMs of each feature for every patient, which is phase-by-phase corresponding to the patient's input 4DCT phase images. Meanwhile, the phase-averaged FM was also calculated by taking an average for ten phase-specific FMs. The phase-averaged FMs have the same size as the input CT and the reference $V_{NM}$ images, and are also voxel-by-voxel aligned. This essentially simulates motion-blurring during image acquisition, allowing direct spatial comparison with reference $V_{NM}$.

## 2.5 Evaluation

In addition to the qualitative visual inspection of the generated FMs, as shown in Figure 3, three main quantitative evaluation methods were also designed, including (a) inter-phase robustness assessments of phase-specific FMs, (b) direct spatial agreement assessments of phase-averaged FMs with reference $V_{NM}$ images, and (c) correlation assessments of spatially averaged feature values with PFT measurements.

For the inter-phase robustness assessment, each resulting phase-specific FM was spatially aligned with the reference $V_{NM}$ image by first using affine registration [33] to align each 4DCT phase image (on which the FM was calculated) and the SPECT attenuation correction CT image. The resulting affine transformation was then applied to FMs. The voxel-wise SCC within lung contour between reference $V_{NM}$ image and transformed phase-specific FMs were phase-by-phase analyzed to assess the spatial agreement. The one-way random effects, single rater/measurement intraclass correlation coefficient (ICC (1, 1)) [34][35] was calculated with SCC values of 10 phases for each feature's FMs, to evaluate the inter-phase robustness quantitatively.

In Section 2.4, phase-averaged FM was also calculated by taking an average for ten phase-specific FMs. Direct spatial agreement assessment between phase-averaged FMs and reference $V_{NM}$ images were performed including voxel-wise SCC within the lung, as well as Dice overlap analysis for the lowest lung function partition (0-33rd percentile of the image signal, $DSC_{LO}$) and highest lung function partition (>66th percentile of the image signal, $DSC_{HI}$).

Moreover, in order to link the regional lung function (FMs) and clinical global function assessment results (PFT measurements), we calculated the correlation between the spatial-averaged feature values (by computing the mean of phase-averaged FM intensity within the lung ROI) and corresponding $FEV_1/FVC$ ratio obtained by PFT for 13 patients. The $FEV_1/FVC$ ratio represents the proportion of the patient's vital capacity that they are able to expire in the first second of forced expiration ($FEV_1$) to the full, forced vital capacity (FVC), and is generally used in the clinical diagnosis of both pulmonary and airway diseases.

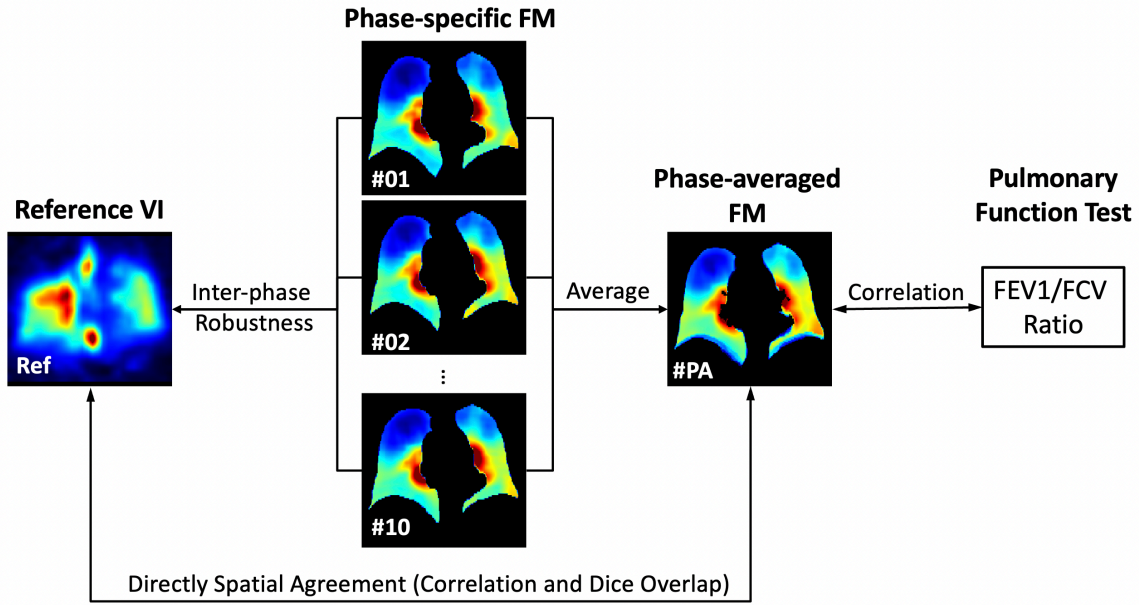

**Figure 3** Schematic for subregional features and FMs quantitative evaluation. VI = ventilation image, FM = feature map, FEV1 = forced expiratory volume in 1 second, FVC = forced vital capacity.

## 3. Results

### 3.1 Feature redundancy

Table 2 summarizes the results of the feature redundancy analysis. The radiomic features clustered in each subset were considered highly correlated, and a total of 16 redundancy subsets were identified.

**Table 2** Results of the feature redundancy analysis.

| Subset # | Feature Name |
|---|---|
| **1** | GLSZM-SizeZoneNonUniformityNormalized, GLSZM-SmallAreaEmphasis |
| **2** | GLSZM-LargeAreaEmphasis, GLSZM-ZoneVariance |
| **3** | GLRLM-LowGrayLevelRunEmphasis, GLRLM-ShortRunLowGrayLevelEmphasis, GLDM-LowGrayLevelEmphasis, GLRLM-LongRunLowGrayLevelEmphasis, GLDM-LargeDependenceLowGrayLevelEmphasis |
| **4** | GLCM-Idn, GLCM-InverseVariance |
| **5** | GLDM-DependenceNonUniformity, GLDM-DependenceNonUniformityNormalized |
| **6** | FirstOrder-Entropy, GLCM-SumEntropy, GLCM-JointEntropy, GLRLM-RunEntropy, GLDM-DependenceEntropy, GLRLM-RunLengthNonUniformityNormalized, GLRLM-RunPercentage, GLRLM-ShortRunEmphasis |

| 7 | GLCM-JointEnergy, GLCM-MaximumProbability, GLRLM-LongRunEmphasis, GLRLM-RunVariance, GLDM-LargeDependenceEmphasis, GLDM-DependenceVariance |
| 8 | GLCM-Id, GLCM-Idm |
| 9 | GLCM-DifferenceAverage, GLCM-DifferenceEntropy, GLSZM-ZonePercentage, GLDM-SmallDependenceEmphasis |
| 10 | First-order-MeanAbsoluteDeviation, First-order-RobustMeanAbsoluteDeviation, First-order-Variance, GLCM-ClusterTendency, GLCM-SumSquares, GLRLM-GrayLevelVariance, GLDM-GrayLevelVariance |
| 11 | GLCM-Autocorrelation, GLCM-SumAverage, GLRLM-HighGrayLevelRunEmphasis, GLRLM-ShortRunHighGrayLevelEmphasis, GLDM-HighGrayLevelEmphasis, GLDM-SmallDependenceHighGrayLevelEmphasis |
| 12 | First-order-Kurtosis, First-order-Skewness, First-order-Uniformity, GLRLM-GrayLevelNonUniformity, GLRLM-GrayLevelNonUniformityNormalized, GLDM-GrayLevelNonUniformity |
| 13 | GLRLM-LongRunHighGrayLevelEmphasis, GLDM-LargeDependenceHighGrayLevelEmphasis |
| 14 | GLSZM-HighGrayLevelZoneEmphasis, GLSZM-SmallAreaHighGrayLevelEmphasis |
| 15 | FirstOrder-Energy, FirstOrder-TotalEnergy, First-order-RootMeanSquared |
| 16 | GLSZM-LowGrayLevelZoneEmphasis, GLSZM-SmallAreaLowGrayLevelEmphasis |

## 3.2 Subregional feature candidate selection

Table 3 lists all the radiomic features with the mean ES greater than 0.330 obtained in section 2.3.3, sorted in descending order by the mean ES. According to the statistical interpretations, an RBC-based ES greater than 0.330 indicates that the feature might have a medium-to-large ability to differentiate defected and non-defected lung CT regions. Combined with the redundancy analysis in section 3.1, the elimination of highly correlated redundant radiomic features left eight first- and second-order feature candidates, including (1) GLSZM Zone Entropy, (2) GLCM Correlation, (3) GLRLM Run Length Non-Uniformity, (4) GLDM Dependence Non-Uniformity, (5) First-order Mean, (6) GLCM Sum Entropy, (7) GLCM Sum Average, and (8) GLCM Imc 2.

**Table 3** Results of subregional feature selection (sorted in descending order by the mean ES). ES = effect size.

| Rank | Feature Name | Mean ES |
|------|--------------|---------|
| 1 | GLSZM - ZoneEntropy | 0.569 |
| 2 | GLCM - Correlation | 0.475 |
| 3 | GLRLM - RunLengthNonUniformity | 0.453 |
| 4 | GLDM - DependenceNonUniformity | 0.450 |
| 5 | First-order - Mean | 0.387 |

| 6 | GLCM - SumEntropy | 0.374 |
|---|---|---|
| 7 | GLCM - SumAverage | 0.370 |
| 8 | GLCM - Autocorrelation | 0.369 |
| 9 | GLCM - JointEntropy | 0.366 |
| 10 | GLCM - Imc2 | 0.361 |
| 11 | GLDM - DependenceNonUniformityNormalized | 0.339 |

## 3.3 FM visual inspection

As the illustrative example, one patient's FMs of previous-selected eight feature candidates are shown in Figure 4. In addition, the lower right shows the superimposed time-average 4DCT image and the corresponding reference $V_{NM}$ ventilation image. On the reference image, a major ventilation defect could be observed in the right upper lobe. The low-signal areas of most FMs show reasonable visual agreement with the reference image in the defect regions, especially for GLDM Dependence Non-Uniformity, GLCM Sum Average, and First-order Mean.

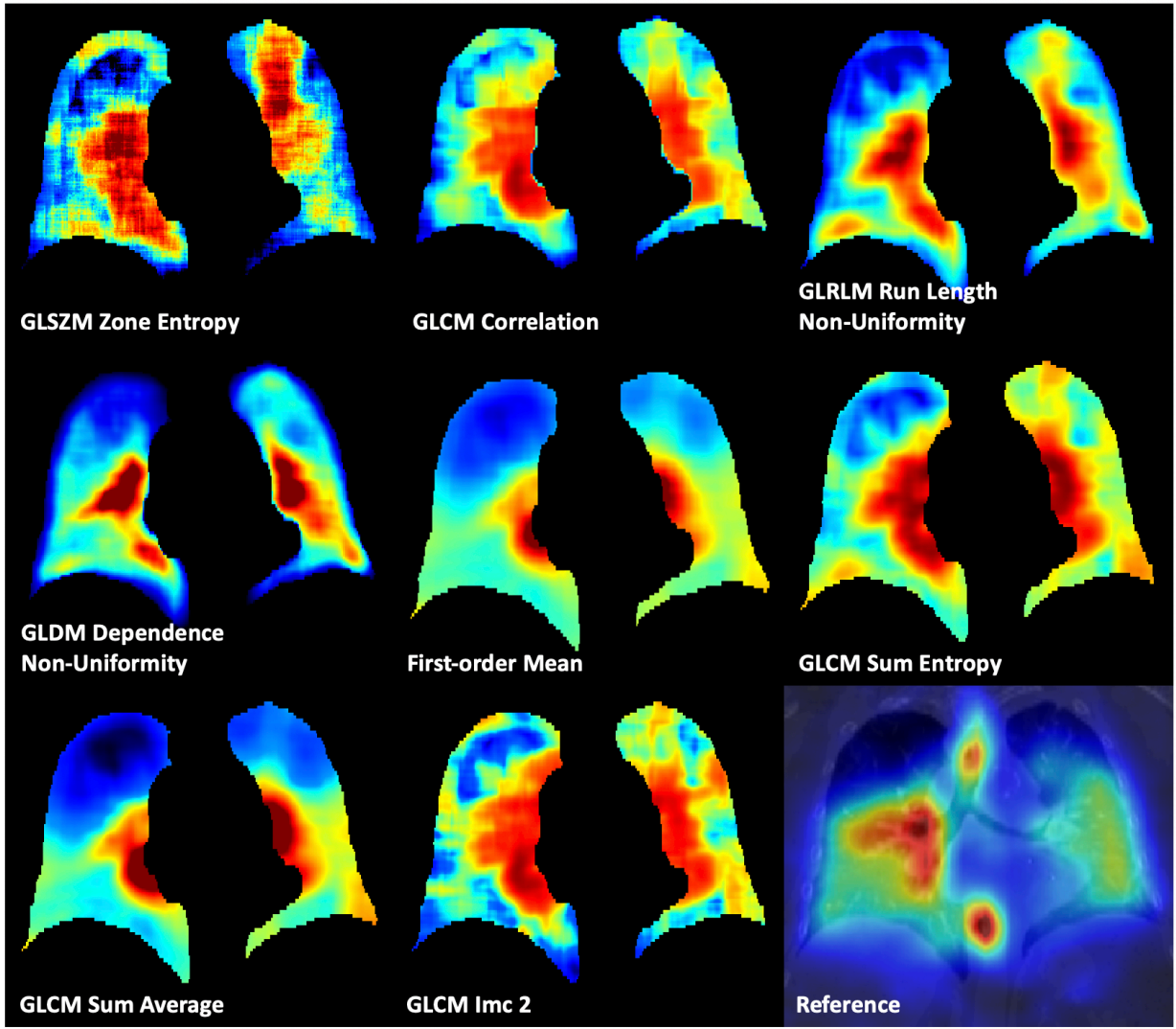

**Figure 4** Visual inspections of reference ventilation image and corresponding resulting FMs of a patient.

### 3.4 Spatial agreement and inter-phase robustness

The box-and-whisker plots in Figure 5 visualize the spatial correlation between FMs

(i.e., the transformed phase-specific FMs and phase-averaged FMs) and the corresponding reference $V_{NM}$. The upper, middle, and lower lines for each box indicate the third quartile, median, and first quartile of 21 patients' SCC values. The black-color cross in each box indicates the mean SCC value. Outliers greater than 1.5 interquartile range (IQR) above the third quartile, or smaller than 1.5 IQR below the first quartile, are also individually plotted as red crosses beyond the whiskers. The quantitative assessment results of the inter-phase robustness, as well as the direct spatial agreement for FMs are additionally summarized in Table 4.

The inter-phase robustness assessment analyzed FM's median SCC with $V_{NM}$ of each feature throughout ten 4DCT breathing phases, and also calculated ICC as the quantitative index to evaluate the FM's inter-phase robustness. We observed that the FMs of these feature candidates showed a relatively robust SCC with $V_{NM}$ across phases. Based on the phase-specific FMs' median SCC range, and the median SCC between phase-averaged FM and reference $V_{NM}$, the highest correlation with reference was found to be the feature GLDM Dependence Non-Uniformity. Specifically, the median SCCs of GLDM Dependence Non-uniformity's phase-specific FMs ranged from 0.54 to 0.59. For direct spatial agreement, across all the 21 patients, its phase-averaged FM achieved the median SCC of 0.60, the median DSC of 0.65 and 0.60 for the lowest and highest lung function partitions.

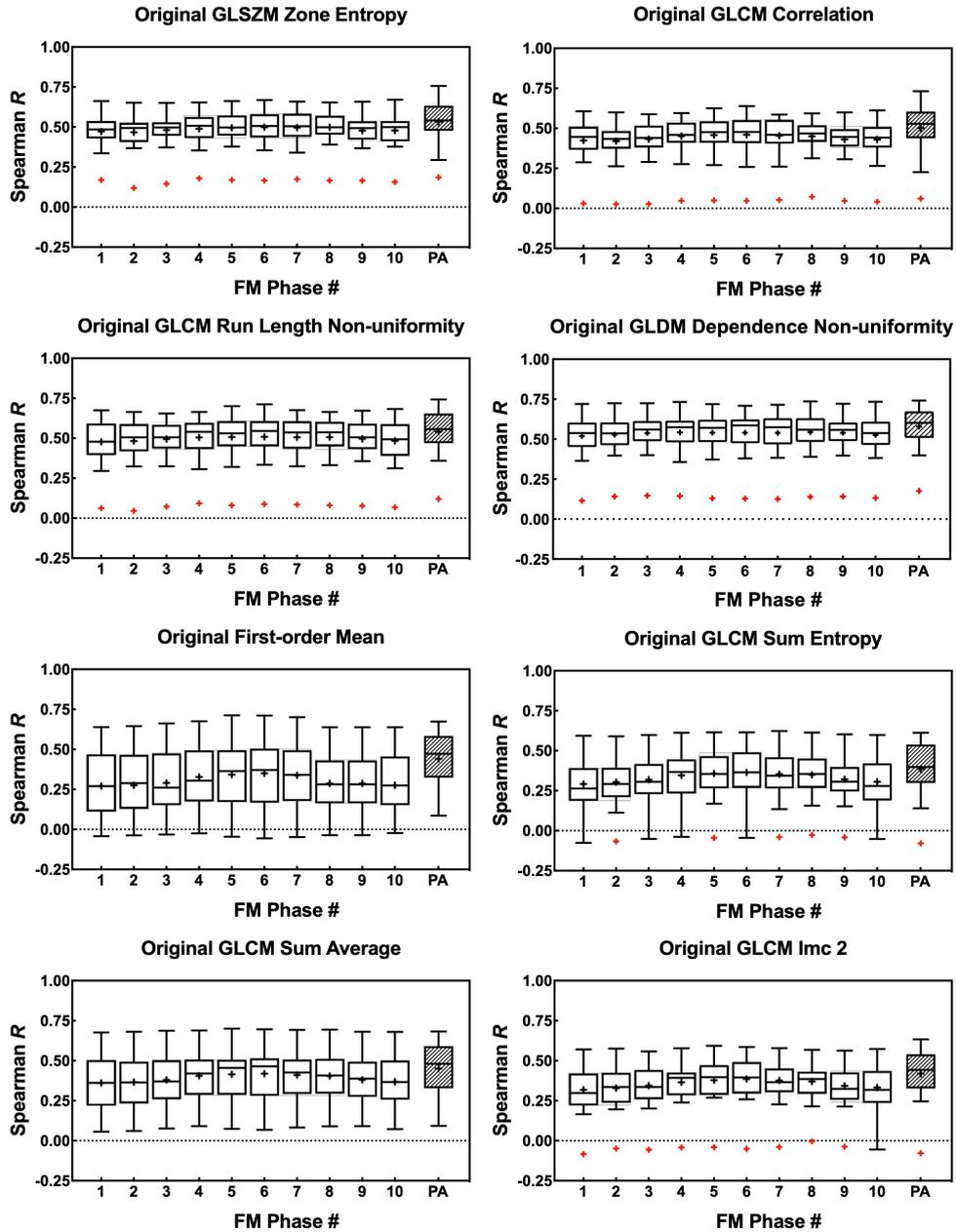

**Figure 5** Spatial correlations between FMs (i.e., the transformed phase-specific FMs and phase-averaged FMs) and corresponding reference ventilation. PA = phase-averaged

**Table 4** Quantitative assessment results of inter-phase robustness and direct spatial agreement for FMs. Ref = reference ventilation image, SCC = Spearman correlation coefficient, IDD = intraclass correlation coefficient, CI = confidence interval, DSC = Dice similarity coefficient.

| Feature Name | Inter-phase Robustness | Direct Spatial Agreement |
|---|---|---|
| | (Ten phase-specific FMs vs. Ref) | (Phase-averaged FM vs. Ref) |

| | Median SCC (Range) | ICC | ICC (95% CI) | Median SCC | Median DSC$_{LO}$ | Median DSC$_{HI}$ |
|---|---|---|---|---|---|---|
| GLSZM Zone Entropy | 0.48 - 0.51 | 0.93 | 0.88 - 0.97 | 0.54 | 0.61 | 0.60 |
| GLCM Correlation | 0.43 - 0.48 | 0.93 | 0.88 - 0.97 | 0.53 | 0.57 | 0.59 |
| GLRLM Run Length Non-Uniformity | 0.48 - 0.55 | 0.96 | 0.94 - 0.98 | 0.56 | 0.62 | 0.60 |
| GLDM Dependence Non-Uniformity | 0.54 - 0.59 | 0.96 | 0.92 - 0.98 | 0.60 | 0.65 | 0.60 |
| GLCM Sum Average | 0.39 - 0.46 | 0.96 | 0.93 - 0.98 | 0.48 | 0.55 | 0.57 |
| GLCM Sum Entropy | 0.26 - 0.37 | 0.94 | 0.89 - 0.97 | 0.40 | 0.50 | 0.52 |
| First-order Mean | 0.26 - 0.37 | 0.95 | 0.90 - 0.98 | 0.47 | 0.54 | 0.56 |
| GLCM Imc 2 | 0.30 - 0.40 | 0.92 | 0.85 - 0.96 | 0.44 | 0.52 | 0.53 |

## 3.5 Correlation with PFT measurements

Scatters and the best-fit lines in Figure 6 provide the relationships between each feature candidate's spatial-averaged feature values and corresponding $FEV_1/FVC$ ratio obtained by PFT measurements for 13 patients. The FMs of feature GLDM Dependence Non-Uniformity showed a significant, moderate-to-high correlation with $FEV_1/FVC$ ratio among eight features.

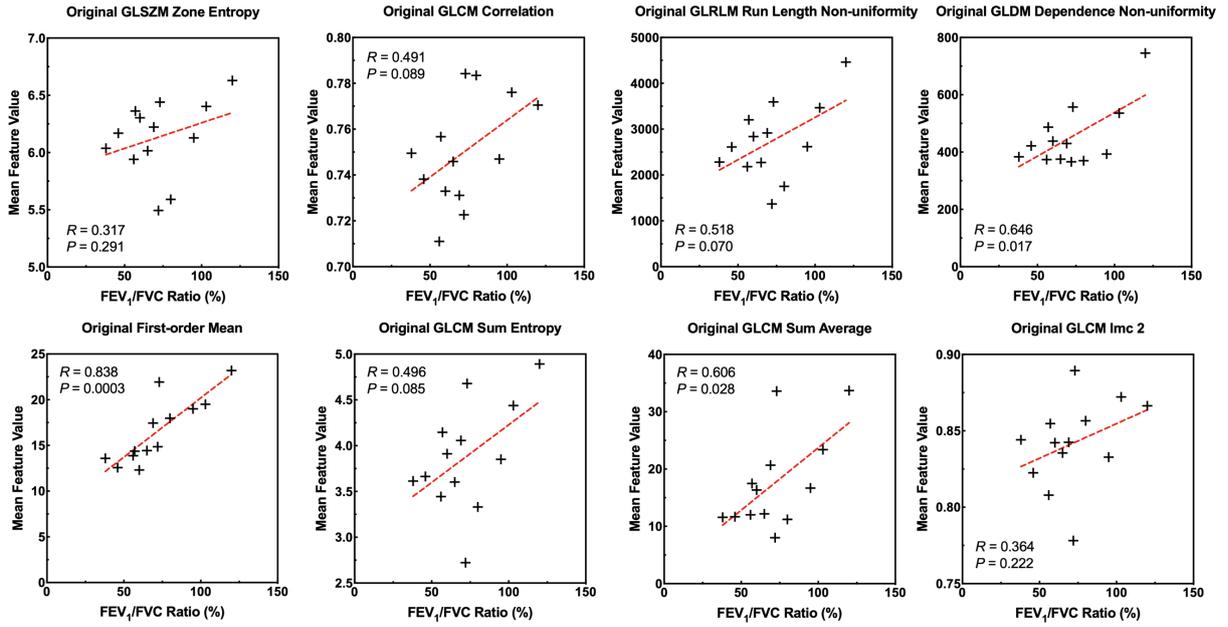

**Figure 6** Correlations between the spatial-averaged feature values (by computing the mean of phase-averaged FM signals of feature candidates within the lung ROI) and corresponding FEV1/FVC ratio obtained by PFT for 13 patients.

## 4. Discussions

The underlying texture information encoded by CT scans was demonstrated in this work to have a certain correlation with both global and local lung function. A sparse-to-fine strategy including subregional feature discovery and voxel-wise feature distribution study, was performed with a cohort of 21 lung cancer RT patients. At the subregion level, through image patch-wise feature extraction, redundancy analysis, and statistics-based feature selection pipeline, we successfully filtered out eight first- and second-order function-correlated radiomic feature candidates without redundancy from a total of 79 features included in this study. These features were observed to have medium-to-large statistical strength to differentiate defected/non-defected lung regions, with statistical ES ranging from 0.339 to 0.569. At the voxel-wise level, FMs of the candidates, including phase-specific FMs and phase-averaged

FMs, were observed to have a moderate-to-strong voxel-wise correlation with reference lung function distributions provided by SPECT ventilation images.

Among eight candidates, the feature Dependence Non-uniformity from GLDM matrix was identified to hold the exciting potential to reflect the lung function heterogeneity encoded by CT texture. Firstly, it achieved first-tier performance in the spatial agreement with NM-based ventilation compared to the published literatures of DIR-based methods according to the systematic review of CTVI (when evaluated with voxel-wise SCC, CTVI algorithms based on two typical DIR-based metrics achieved the average performance of 0.45 and 0.41 respectively) [12]. Secondly, both the exploratory results and the mathematical nature support that the identified feature and it's FM have robustness. In the robustness assessment throughout the breathing cycle, the performance of ICC on ten respiratory phases was 0.96 (95%CI: 0.92 - 0.98), indicating the FM could be excellently reproduced at different inflation level, as shown in Figure 5. The study on phantom and human's lung region [36] also examined the reproducibility of the identified feature across different scanners, acquisition parameters and reconstruction parameters, and the feature GLDM Dependence Non-uniformity was found to be robust against all above changes. The feature is also intrinsically invariant against image translation/rotation and linear transformation/shift of grey level according to the mathematical characteristic of gray level dependence matrix design [38], and thus, the performance of our identified feature could be reproducible and stable under variant image preprocessing workflows. Thirdly, the mapping of feature distribution also offered greater explainability than AI based method. Instead of the black box in the published AI-based methods [17][18][19], our method provides a definitive mathematical algorithm with a sliding window technique for function distribution mapping. From the mathematical perspective, the values of GLDM Dependence Non-uniformity measure the homogeneity among dependencies of the image intensity, which are proportionally related to the coarseness of the image [37]. Our results, therefore, suggest that the coarseness of CT local texture is correlated with the local lung function.

The mapping of feature distribution in this study adopted a novel angle for CTVI generation after demonstrating that CT-encoded textures are correlated with local lung function. Conventionally, the DIR-based CTVI were synthesized by assuming the association between lung ventilation and breathing-induced surrogate changes, which can be described via performing DIR to recover inter-respiratory-phase lung motions. The accuracy of the synthetic function distribution via DIR methods is sensitive to DIR algorithms and parameters, and it also requires the presence of breathing-correlated images (e.g., 4DCT or BHCT scans). The texture-based method, on the other hand, is robust under various conditions, such as different respiratory phases and randomness, and it only requires a single static scan to map the lung function. The DIR- and texture-based methods attempt to assess the CTVI from different angles, and therefore, it is likely that these two methods offer different functional information and a fusion between both methods could yield a comprehensive representation of lung function from CT.

Technically, our research framework facilitates the further identification of function-correlated radiomic features by implementing a sparse-to-fine strategy. In the feature discovery stage, the subsampling of lung CT images into compact and quasi-equally sized patches before extracting subregional radiomic features is due to two considerations. Firstly, by limiting the calculation area of radiomic features to patches with equal volumes, it was ensured that the derived values were determined entirely by the regional texture information, avoiding that some features were incorrectly considered to have defected/non-defected lung regions differentiating capabilities due to the volume differences of the extracted area [39]. Secondly, by deliberately setting the volume of generated patches to be identical to the volume of the sliding window used in the subsequently voxel-wise FM calculation ($21 \times 21 \times 21 = 9261mm^3$), we expect the feature candidates thus filtered out to have inherited differentiating capabilities at both subregion level and voxel-wise level. Compared to extracting subregional feature values from divided patches of CT images ($\sim 1 \times 10^2$ patches per patient), the computation of 3D voxel-wise FMs is a computationally intensive and time-consuming task ($\sim 1 \times 10^6$ voxels per patient). Through the sparse-to-fine strategy, only limited number of potential function-correlated feature candidates were identified prior to FMs generations and feature analysis. This will significantly reduce the cost of target feature discovery, and increase the flexibility of adjusting the feature categories covered in the study. Based on the current framework, more feature types can easily be taken into account in the subsequent research, including first-order and second-order features from the Laplacian of Gaussian (LoG) and the wavelet filtered image domains.

This research still leaves plenty of room for improvement. Before subregional feature extraction and voxel-wise FM calculation tasks, we discretized the grey levels of all the input CT images into 64 bins within [-1000, 0] HU. Grey level discretization is a critical part of quantitative radiomic studies, which could reduce the infinite possible number of intensity values from the raw image to a finite set, and effectively suppress the effects of noise [40][41].

As expected, the various discretization methods will impact the calculating results and repeatability of the radiomic feature values. Since there are currently no specific guidelines as to what constitutes an optimal re-binning, it will be necessary to further explore the effects of discretization for each feature category to improve the FM performance and make the calculation more reliable and reproducible. Similarly, the size of the sliding window for FM generation, which determined how many voxels would be included in the texture quantification, is another critical factor that needs further optimization in the following study. The setting of sliding window size for FM generation ($21 \times 21 \times 21mm^3$) was primarily based on lung physiological considerations in this work. During feature distribution mapping, each sliding window was expected to cover multiple lung minimum gas exchange units with a measured size slightly larger than $10 \times 10 \times 10mm^3$ [42], due to the demonstrated tendency of lung function to be spatially clustered [42][43]. A too small sliding window size might make the calculated FM suffer from noise, while a too large size will reduce function-correlated information capture sensitivity and significantly increase computational complexity. Finally, in addition to abovementioned technical improvements, the more standardized implementation protocol, and external validations on method's generalizability are also required.

## 5. Conclusion

In this work, we extracted and identified underlying lung function-correlated textures, and studied their spatial distribution and robustness from thoracic CT scans through a sparse-to-fine radiomics framework. Among the feature categories considered so far, the selected feature candidates were observed to have moderate-to-strong correlation with both global and local lung function. The results have the exciting potential to further the understanding of the underlying relationships between regional function and lung texture information, and to facilitate more accurate lung disease diagnosis.

## 6. Acknowledgments

The authors thank Dr. Tokihiro Yamamoto (University of California Davis School of Medicine) and Dr. John Kipritidis (Northern Sydney Cancer Centre) for supporting this work with unabridged acquisition information and spirometry measurements, as these data were not fully available in the raw VAMPIRE challenge dataset.

## Disclosure of Funding

This work was supported in part by (1) General Research Fund (GRF 15103520) from the University Grants Committee, (2) Health and Medical Research Fund (HMRF 07183266) from the Food and Health Bureau, The Government of the Hong Kong Special Administrative Regions, and (3) Shenzhen-Hong Kong-Macau S&T Program (Category C) (SGDX20201103095002019) of the Shenzhen Science and Technology Innovation Committee.

## Conflict of Interest Statement

The authors declare that they have no known competing financial interests or personal relationships that could have appeared to influence the work reported in this paper.

## Data Sharing Statements

Research data is not available at this time.

## 7. Reference

[1] Glenny, R. W., & Robertson, H. T. (2011). Spatial distribution of ventilation and perfusion: Mechanisms and regulation. Comprehensive Physiology, 1(1), 373–395.

[2] Barrow, A., & Pandit, J. J. (2017). Lung ventilation and the physiology of breathing. Surgery (Oxford), 35(5), 227–233.

[3] Simon, B. A., Kaczka, D. W., Bankier, A. A., & Parraga, G. (2012). What can computed tomography and magnetic resonance imaging tell us about ventilation? Journal of Applied Physiology, 113(4), 647–657.

[4] Yamamoto, T., Kabus, S., Von Berg, J., Lorenz, C., & Keall, P. J. (2011). Impact of four-dimensional computed tomography pulmonary ventilation imaging-based functional avoidance for lung cancer radiotherapy. International Journal of Radiation Oncology Biology Physics, 79(1), 279–288.

[5] Ren, G., Zhang, J., Li, T., Xiao, H., Cheung, L. Y., Ho, W. Y., … Cai, J. (2021). Deep Learning-Based Computed Tomography Perfusion Mapping (DL-CTPM) for Pulmonary CT-to-Perfusion Translation. International Journal of Radiation Oncology*Biology*Physics.

[6] Eslick, E.M., M.J. Stevens, and D.L. Bailey (2019). SPECT V/Q in Lung Cancer Radiotherapy Planning. Seminars in Nuclear Medicine, 49(1), 31-36.

[7] Le Roux, P.Y., et al.. (2019). PET/CT Lung Ventilation and Perfusion Scanning using Galligas and Gallium-68-MAA. Seminars in Nuclear Medicine, 49(1), 71-81.

[8] Tahir, B. A., Van Holsbeke, C., Ireland, R. H., Swift, A. J., Horn, F. C., Marshall, H., … Wild, J. M. (2016). Comparison of CT-based lobar ventilation with 3He MR imaging

ventilation measurements. Radiology, 278(2), 585–592.

[9] Tahir, B. A., Marshall, H., Hughes, P. J. C., Brightling, C. E., Collier, G., Ireland, R. H., & Wild, J. M. (2019). Comparison of CT ventilation imaging and hyperpolarised gas MRI: Effects of breathing manoeuvre. Physics in Medicine and Biology, 64(5).

[10] Eichinger, M., Puderbach, M., Fink, C., Gahr, J., Ley, S., Plathow, C., … Kauczor, H. U. (2006). Contrast-enhanced 3D MRI of lung perfusion in children with cystic fibrosis - Initial results. European Radiology, 16(10), 2147–2152.

[11] Kipritidis, J., Tahir, B. A., Cazoulat, G., Hofman, M. S., Siva, S., Callahan, J., … Patton, T. J. (2019). The VAMPIRE challenge: A multi-institutional validation study of CT ventilation imaging. Medical Physics, 49(3), 1198–1217.

[12] Hegi-Johnson, F., de Ruysscher, D., Keall, P., Hendriks, L., Vinogradskiy, Y., Yamamoto, T., … Kipritidis, J. (2019). Imaging of regional ventilation: Is CT ventilation imaging the answer? A systematic review of the validation data. Radiotherapy and Oncology, 137, 175–185.

[13] Cai, J., Altes, T. A., Miller, G. W., Sheng, K., Read, P. W., Mata, J. F., … Brookeman, J. R. (2007). MR grid-tagging using hyperpolarized helium-3 for regional quantitative assessment of pulmonary biomechanics and ventilation. Magnetic Resonance in Medicine, 58(2), 373–380.

[14] Cai, J., Miller, G. W., Altes, T. A., Read, P. W., Benedict, S. H., de Lange, E. E., … Sheng, K. (2007). Direct Measurement of Lung Motion Using Hyperpolarized Helium-3 MR Tagging. International Journal of Radiation Oncology Biology Physics, 68(3), 650–653.

[15] Cai, J., Sheng, K., Benedict, S. H., Read, P. W., Larner, J. M., Mugler, J. P., … Miller, G.

W. (2009). Dynamic MRI of Grid-Tagged Hyperpolarized Helium-3 for the Assessment of Lung Motion During Breathing. International Journal of Radiation Oncology Biology Physics, 75(1), 276–284.

[16] Bauman, G., Puderbach, M., Deimling, M., Jellus, V., Chefd'hotel, C., Dinkel, J., … Schad, L. R. (2009). Non-contrast-enhanced perfusion and ventilation assessment of the human lung by means of fourier decomposition in proton MRI. Magnetic Resonance in Medicine, 62(3), 656–664.

[17] Zhong, Y., Vinogradskiy, Y., Chen, L., Myziuk, N., Castillo, R., Castillo, E., … Wang, J. (2019). Technical Note: Deriving ventilation imaging from 4DCT by deep convolutional neural network. Medical Physics, 46(5), 2323–2329.

[18] Liu, Z., Miao, J., Huang, P., Wang, W., Wang, X., Zhai, Y., … Dai, J. (2020). A deep learning method for producing ventilation images from 4DCT: First comparison with technegas SPECT ventilation. Medical Physics, 47(3), 1249–1257.

[19] Grover, J., Byrne, H. L., Sun, Y., Kipritidis, J., & Keall, P. (2022). Investigating the use of machine learning to generate ventilation images from CT scans. Medical Physics, 49(8), 5258–5267.

[20] Van Griethuysen, J. J. M., Fedorov, A., Parmar, C., Hosny, A., Aucoin, N., Narayan, V., … Aerts, H. J. W. L. (2017). Computational radiomics system to decode the radiographic phenotype. Cancer Research, 77(21), e104–e107.

[21] Frix, A. N., Cousin, F., Refaee, T., Bottari, F., Vaidyanathan, A., Desir, C., … Guiot, J. (2021). Radiomics in lung diseases imaging: State-of-the-art for clinicians. Journal of Personalized Medicine, 11(7).

[22] Lafata, K. J., Zhou, Z., Liu, J. G., Hong, J., Kelsey, C. R., & Yin, F. F. (2019). An Exploratory Radiomics Approach to Quantifying Pulmonary Function in CT Images. Scientific Reports, 9(1), 1–9.

[23] Westcott, A., Capaldi, D. P. I., McCormack, D. G., Ward, A. D., Fenster, A., & Parraga, G. (2019). Chronic obstructive pulmonary disease: Thoracic CT texture analysis and machine learning to predict pulmonary ventilation. Radiology, 293(3), 676–684.

[24] Yamamoto, T., Kabus, S., Lorenz, C., Mittra, E., Hong, J. C., Chung, M., … Keall, P. J. (2014). Pulmonary ventilation imaging based on 4-dimensional computed tomography: Comparison with pulmonary function tests and SPECT ventilation images. International Journal of Radiation Oncology Biology Physics, 90(2), 414–422.

[25] Achanta, R., Shaji, A., Smith, K., Lucchi, A., Fua, P., &Süsstrunk, S. (2012). SLIC Superpixels Compared to State-of-the-Art Superpixel Methods. IEEE Transactions on Pattern Analysis and Machine Intelligence, 34(11), 2274–2282.

[26] Irving, B. (2016). maskSLIC: Regional Superpixel Generation with Application to Local Pathology Characterisation in Medical Images. 1–7.

[27] Faught, A. M., Miyasaka, Y., Kadoya, N., Castillo, R., Castillo, E., Vinogradskiy, Y., &Yamamoto, T. (2017). Evaluating the Toxicity Reduction With Computed Tomographic Ventilation Functional Avoidance Radiation Therapy. International Journal of Radiation Oncology Biology Physics, 99(2), 325–333.

[28] Waxweiler, T. V., Schubert, L. K., Diot, Q., Castillo, R., Castillo, E., Guerrero, T. M., Gaspar, L. E., Miften, M., Kavanagh, B. D., &Vinogradskiy, Y. (2015). Towards a 4DCT-

Ventilation Functional Avoidance Clinical Trial: Determining Patient Eligibility. *International Journal of Radiation Oncology\*Biology\*Physics*, *93*(3), E416–E417.

[29] Ester, M., Kriegel, H.-P., Sander, J., & Xu, X. (1996). A density-based algorithm for discovering clusters in large spatial databases with noise. In Proceedings of the 2nd International Conference on Knowledge Discovery and Data Mining (pp. 226–231). Portland: AAAI Press.

[30] Cureton, E. E. (1956). Rank-biserial correlation. Psychometrika, 21(3), 287-290.

[31] Kerby, D. S. (2014). The simple difference formula: An approach to teaching nonparametric correlation. Comprehensive Psychology, 3, 11-IT.

[32] Cliff, N. (1993). Dominance statistics: Ordinal analyses to answer ordinal questions. Psychological bulletin, 114(3), 494.

[33] Sandkühler, R., Jud, C., Andermatt, S., &Cattin, P. C. (2018). AirLab: Autograd Image Registration Laboratory. http://arxiv.org/abs/1806.09907

[34] Shrout, P. E., & Fleiss, J. L. (1979). Intraclass correlations: Uses in assessing rater reliability. Psychological Bulletin, 86(2), 420–428.

[35] Koo, T. K., & Li, M. Y. (2016). A Guideline of Selecting and Reporting Intraclass Correlation Coefficients for Reliability Research. Journal of Chiropractic Medicine, 15(2), 155–163.

[36] Jha, A. K., Mithun, S., Jaiswar, V., Sherkhane, U. B., Purandare, N. C., Prabhash, K., … Traverso, A. (2021). Repeatability and reproducibility study of radiomic features on a phantom and human cohort. Scientific Reports, 11(1), 1–12.

[37] Hu, J., Zhao, Y., Li, M., Liu, J., Wang, F., Weng, Q., … Cao, D. (2020). Machine learning-based radiomics analysis in predicting the meningioma grade using multiparametric MRI.

European Journal of Radiology, 131(July), 109251.

[38] Sun, C., & Wee, W. G. (1983). Neighboring gray level dependence matrix for texture classification. Computer Vision, Graphics and Image Processing, 23(3), 341–352.

[39] Welch, M. L., McIntosh, C., Haibe-Kains, B., Milosevic, M. F., Wee, L., Dekker, A., … Jaffray, D. A. (2019). Vulnerabilities of radiomic signature development: The need for safeguards. Radiotherapy and Oncology, 130, 2–9.

[40] Leijenaar, R. T. H., Nalbantov, G., Carvalho, S., Van Elmpt, W. J. C., Troost, E. G. C., Boellaard, R., … Lambin, P. (2015). The effect of SUV discretization in quantitative FDG-PET Radiomics: The need for standardized methodology in tumor texture analysis.

[41] Tixier, F., Le Rest, C. C., Hatt, M., Albarghach, N., Pradier, O., Metges, J. P., … Visvikis, D. (2011). Intratumor heterogeneity characterized by textural features on baseline 18F-FDG PET images predicts response to concomitant radiochemotherapy in esophageal cancer. Journal of Nuclear Medicine, 52(3), 369–378.

[42] Levin, D. L., Schiebler, M. L., & Hopkins, S. R. (2017). Physiology for the pulmonary functional imager. European Journal of Radiology, 86, 308–312.

[43] Tahir, B. A., Hughes, P. J. C., Robinson, S. D., Marshall, H., Stewart, N. J., Norquay, G., … Ireland, R. H. (2018). Spatial Comparison of CT-Based Surrogates of Lung Ventilation With Hyperpolarized Helium-3 and Xenon-129 Gas MRI in Patients Undergoing Radiation Therapy. International Journal of Radiation Oncology Biology Physics, 102(4), 1276–1286.